\documentclass[%
 reprint,
 amsmath,amssymb,
 aps,
superscriptaddress
]{revtex4-2}

\usepackage[dvipsnames]{xcolor}
\usepackage{tabularx}
\usepackage{graphicx}
\usepackage{dcolumn}
\usepackage{bm}
\usepackage{cleveref}
\usepackage{multirow}

\begin{document}


\title{Physics-based modeling of cyclic and calendar aging of LIBs with Si-Gr composite anodes} 

\author{Micha C. J.~Philipp}
 \affiliation{Institute for Engineering Thermodynamics, German Aerospace Center (DLR), Wilhelm-Runge-Straße 10, 89081 Ulm, Germany}
 \affiliation{Helmholtz Institute Ulm (HIU), Helmholtzstraße 11, 89081 Ulm, Germany}
 \affiliation{Faculty of Natural Sciences, Ulm University, Albert-Einstein-Allee 11, 89081 Ulm, Germany}
\author{Lukas Köbbing}
 \affiliation{Institute for Engineering Thermodynamics, German Aerospace Center (DLR), Wilhelm-Runge-Straße 10, 89081 Ulm, Germany}
 \affiliation{Helmholtz Institute Ulm (HIU), Helmholtzstraße 11, 89081 Ulm, Germany}
 \affiliation{Faculty of Natural Sciences, Ulm University, Albert-Einstein-Allee 11, 89081 Ulm, Germany}
\author{Alexander Karger}
\affiliation{Technical University of Munich, School of Engineering and Design, Department of Energy and Process Engineering, Institute for Electrical Energy Storage
Technology, Arcissstr. 21, 80333 Munich, Germany}
\author{Andreas Jossen}
\affiliation{Technical University of Munich, School of Engineering and Design, Department of Energy and Process Engineering, Institute for Electrical Energy Storage
Technology, Arcissstr. 21, 80333 Munich, Germany}
\author{Arnulf Latz}
 \affiliation{Institute for Engineering Thermodynamics, German Aerospace Center (DLR), Wilhelm-Runge-Straße 10, 89081 Ulm, Germany}
 \affiliation{Helmholtz Institute Ulm (HIU), Helmholtzstraße 11, 89081 Ulm, Germany}
 \affiliation{Faculty of Natural Sciences, Ulm University, Albert-Einstein-Allee 11, 89081 Ulm, Germany}
\author{Birger Horstmann}
 \email{birger.horstmann@dlr.de}
 \affiliation{Institute for Engineering Thermodynamics, German Aerospace Center (DLR), Wilhelm-Runge-Straße 10, 89081 Ulm, Germany}
 \affiliation{Helmholtz Institute Ulm (HIU), Helmholtzstraße 11, 89081 Ulm, Germany}
 \affiliation{Faculty of Natural Sciences, Ulm University, Albert-Einstein-Allee 11, 89081 Ulm, Germany}

\date{\today}

\begin{abstract}
Higher energy density and longer lifetime are the requirements for next-generation lithium-ion batteries. A promising anode material is silicon, which offers high specific capacity, but its significant volume change during lithiation and delithiation enormously reduces battery lifetime. A physical understanding of the processes degrading the battery is key to mitigate this effect and advance in the field. We develop a physics-based model to describe degradation during battery cycling under various protocols and storage conditions, with varying check-up (CU) frequencies. The model can disentangle basic degradation mechanisms, such as the growth of the Solid-Electrolyte Interphase (SEI), from silicon mechanisms, such as particle cracking, SEI growth on cracks, and loss of active material (LAM). We investigate the impact of CUs on the observed storage degradation and the reason behind the increased degradation in batteries, including silicon in the anode. Additionally, we relate the observed degradation to operating conditions, enabling future optimization of battery use and design.

\begin{description}
\item[keywords]
SEI growth --- particle cracking --- loss of active material --- battery degradation --- silicon graphite anode
\end{description}
\end{abstract}

\maketitle



\section{Introduction}

Longer lifetimes and higher gravimetric energy densities are needed, to make Lithium-ion batteries (LIB) even more attractable for electric vehicles (EV). A strategy to achieve higher energy densities is to include materials with high specific capacity in the negative electrode. Silicon is such a material, but it comes along with a substantial impact on battery degradation. Therefore, a physical understanding of battery degradation and the influence of silicon is indispensable for prolonging battery lifetime and advancing electrification.

Battery degradation comprises a variety of complex coupled mechanisms \cite{Vetter2005, Birkl2017, Edge2021, Barr2013}. The occurrence and strength of these mechanisms depend on battery materials and usage profiles. In principle, the key mechanisms are known, but their relative impact on battery capacity fade, the timing of their occurrence, and their dependence on operating conditions remain vague. The main degradation modes occurring in LIBs are loss of lithium inventory (LLI), loss of active material (LAM), and loss of electrolyte (LE). The underlying mechanisms considered responsible and lifetime-limiting are SEI growth \cite{Peled2017, Winter2009} and lithium plating \cite{Waldmann2018, Liu2016, Campbell2019} for LLI, particle cracking for LAM \cite{Reniers2019, Edge2021}, and electrolyte dry-out \cite{Liao2024} for LE and LAM. The strong coupling between these degradation mechanisms \cite{OKane2022} and the lack of individual markers for each mechanism render the essential task of accurately modeling the mechanisms extremely challenging. Therefore, isolating the degradation mechanisms to obtain valid physical conclusions and models about the individual processes is crucial.

The dependence on operating conditions can be used to isolate the mechanisms \cite{Vetter2005}. For example, at high temperatures and low charging currents, Li-plating becomes negligible. Then, SEI growth is the dominating degradation mechanism for LLI \cite{Keil2016}. It can be further isolated from other degradation mechanisms by investigating storage experiments. \citet{Single2018} analyzed capacity fade under long-term storage conditions to find a valid SEI growth model. As a result, electron diffusion was identified as a highly promising growth mechanism \cite{Köbbing2023, Philipp2025}. However, questions like the impact of performing check-ups (CU) during the experiment on the capacity fade and the transferability of electron diffusion to SEI growth during battery cycling remained unanswered. With a reliable SEI growth model, the other contributions to battery degradation become more accessible and identifiable.

The usage of silicon in the negative electrode contributes significantly to the degradation \cite{Majherova2025, Li2019, Su2014}. Especially the substantial volume change of silicon during de-/lithiation can lead to loss of contact and cracks in the silicon particles and enveloping SEI layers. This may result in a high LAM of silicon ($\textrm{LAM}_\textrm{Si}$) and increased LLI due to SEI healing. The consequences, such as changes in active surface area due to $\textrm{LAM}_\textrm{Si}$, make degradation with silicon more complex, but necessary to capture the basic understanding. In previous work, cracking and the amount of LAM have been related to stresses in silicon particles \cite{Deshpande2012, Laresgoiti2015, Bonkile2024}. This relation is based on empirical findings of material fatigue due to repeated mechanical loading \cite{Basquin1910}. However, the transferability of this relation to silicon in the negative electrode, the evolution of cracks over many cycles, and the contribution to capacity loss (CL) are not accurately assessed.

In this work, we develop a physics-based model that incorporates SEI growth and silicon cracking to describe the observed aging during cycling and storage, including CUs. We investigate which degradation mechanisms contribute most to capacity fade under different operating conditions. By doing so, we analyze the importance of accounting for the effect of CUs in modeling storage aging. In Sec. \ref{sec:Methods}, we describe the experimental setups, the battery model, and the applied degradation models. In Sec. \ref{sec:Results}, we compare our model results to experimental findings and discuss our findings along the way. We conclude in Sec. \ref{sec:Conclusion}.


\begin{figure*}[]
	\centering
	\makebox[0pt]{%
    \includegraphics[width=0.85\paperwidth]{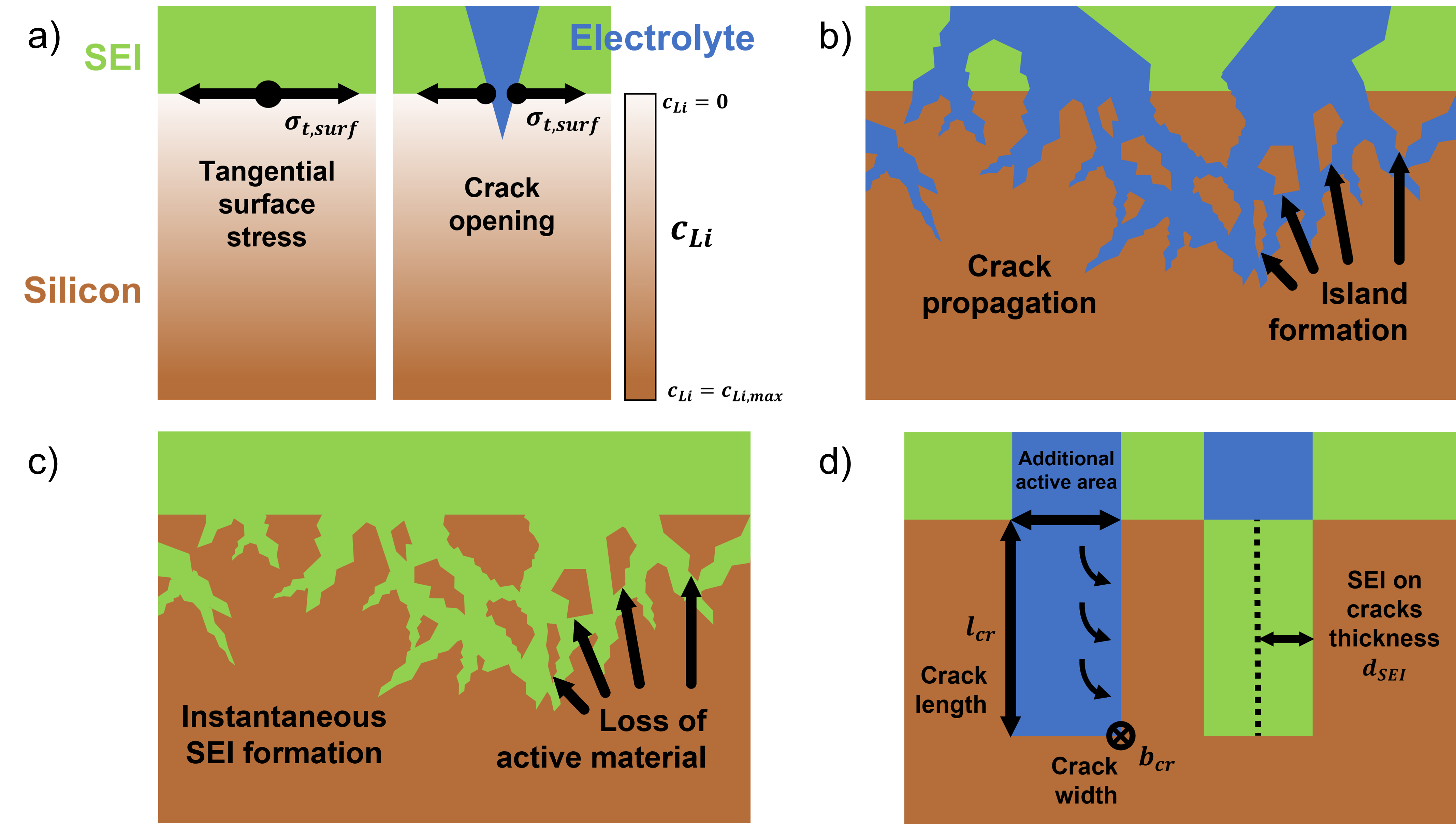}}
	\caption{Schematic illustration of the degradation mechanisms occurring on silicon: a) continuous SEI growth on the active surface area and tangential stress due to a Li-ion concentration gradient that causes crack opening or growth, b) crack and electrolyte propagation into the particle by battery cycling and consequential silicon island formation, c) instantaneous SEI growth on the surface exposed to electrolyte and $\textrm{LAM}_\textrm{Si}$, d) effective crack description for the model: the crack length $l_\mathrm{cr}$ and crack width $b_\mathrm{cr}$ (perpendicular to the illustration) give the area exposed to electrolyte due to cracking, a fraction of the crack-area as additional active area used for Li-ions to intercalate into the silicon particle upon which also continuous SEI is growing, the thickness $d_\mathrm{SEI}$ of the instantaneously grown SEI after crack formation is filling the whole volume of the crack inside the particle.}
	\label{fig.Crack-Illustration}
\end{figure*}

\section{Methods}\label{sec:Methods}

\subsection{Experimental Data}\label{subsec:ExpData}

The experimental data stems from measurements performed by \citet{Wildfeuer2023} with Sony Murata US18650VTC5A battery cells. The cells contain a silicon-graphite (SiGr) composite anode (1.4\% wt Si \cite{Lain2019}) and an NCA cathode for which the qOCVs were measured. For further cell details or the qOCVs, we refer to the previous publications \cite{Wildfeuer2023, Schmitt2022, Kitada2019}. \citet{Wildfeuer2023} performed storage and cycling experiments at various test conditions. These experiments have been analyzed with mechanistic modeling of storage and cycling aging \cite{Karger2023, Karger2024}, as well as empirical modeling of particle versus SEI cracking \cite{Karger2024Crack}. Complementing, we present a physics-based approach in this work covering all of these aspects within one consistent model. In the following, we summarize the important characteristics of these experiments.

In the storage experiments, the cells have been stored for 96 weeks in total at different state-of-charge (SoC) and temperature conditions. During the experiment, CUs were performed to assess the degradation over time. Additionally, \citet{Wildfeuer2023} varied the CU frequency (6/12/96 week interval), yielding different total numbers of performed CUs (16/8/1) at the end of the calendar aging experiment.

Despite the storage experiments, \citet{Wildfeuer2023} also conducted a variety of cycling experiments using a D-optimal design of experiments (see Tab. 1 in \cite{Karger2024}). These experiments differ in their cycling protocols, including charge and discharge rates, depth-of-discharge (DoD), mean SoC, and temperatures. After 50 equivalent full cycles (EFCs), CUs were performed again to assess the degradation over time. Additionally, a cell was repeatedly cycled with the CU-protocol only to investigate the effect and degradation of the CUs themselves. The cycling cells analysed in the main manuscript are listed in Tab. \ref{tab.Cycl_Protocols} with their corresponding protocols. Note that the cell surface temperature measured by individual sensors can deviate considerably from the chamber's target temperature. As the temperature follows a gradient from the heating source to the cooling at the cell surface, we estimate the mean cell temperature to be slightly higher than the measured surface temperature. The estimated mean cell temperature is then used to simulate the specific cell protocol.

The CU procedure is at $T = 20$°C regardless of storage or cycling temperature and is equivalent for each cell, despite the recharging phase to the storage or cycling SoC. The CU consists of an initial resting phase (three hours), a discharge to the lower voltage cut-off, followed by a second rest period (two hours). Afterwards, two full CC-CV cycles with 1C to determine the capacity are executed. Next, the cells are charged with C/10 to obtain the qOCV. This qOCV is used to perform half-cell fitting to extract the state-of-health (SoH) variables (e.g., LLI, SoH of the positive electrode $\mathrm{SoH}_\mathrm{PE}$, SoH of the silicon in the negative electrode $\mathrm{SoH}_\mathrm{Si}$, SoH of the graphite in the negative electrode $\mathrm{SoH}_\mathrm{Gr}$, ...). The SoH variables measure the available capacity in relation to the initial total capacity of the corresponding material ($\mathrm{SoH}_i = \frac{C_i}{C_{i,0}}$) and are connected to the LAM by $\mathrm{SoH}_i = 1- \mathrm{LAM}_i$. After the charging, three discharge pulses (with intermediate resting and 1C discharge phases) at different SoCs (100\%/50\%/20\%) are performed to calculate the cell resistance. Finally, the cell becomes charged or discharged to the target cycling or storage SoC.

For specific storage conditions, \citet{Wildfeuer2023} investigated cell-to-cell variations and found them to be minimal. But most test protocols were performed with only one cell. Hence, the uncertainty is not known for all cases.

As the vast majority of battery processes are temperature-dependent, we focus on the most stable and reliable test conditions ($T = 20$°C for storage and cycling) to minimize thermal effects. For accurate model statements at higher temperatures, the thermal dependencies of all processes have to be determined at a cost, which is beyond the scope of this work. Moreover, additional degradation mechanisms may come into play at higher temperatures. However, we also compare our findings with stable cells at higher temperatures.

\begin{figure*}[]
	\centering
	\makebox[0pt]{%
    \includegraphics[width=0.85\paperwidth]{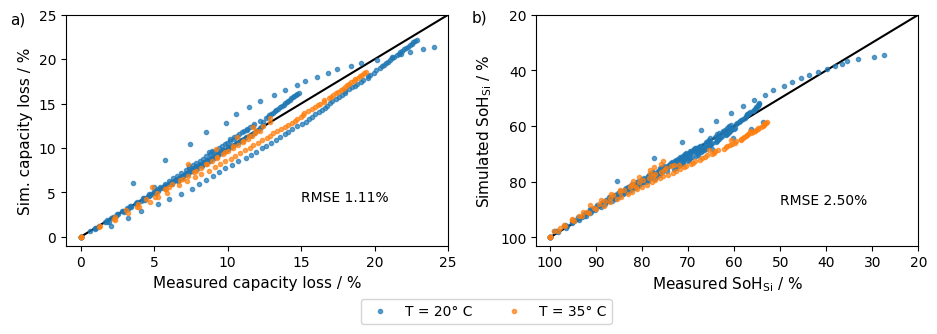}}
	\caption{Simulated versus measured a) CL and b) $\mathrm{SoH}_\mathrm{Si}$ for the cells listed in Tab. \ref{tab.Cycl_Protocols} with different temperatures ($T = 20$°C and $T = 35$°C) and different cycling protocols (C-rates, mean SoC, DoD).}
	\label{fig.Cycl-Summary-CL-SoH}
\end{figure*}

\subsection{Digital Cell}

To simulate the battery behavior, we use a physics-based battery model. This model describes the various physicochemical effects within a battery using a set of coupled partial differential equations (PDEs). For the most accurate description, these PDEs have to be solved numerically, which is costly in 3D. Due to the computational effort, this is infeasible for a whole battery and long time periods. One has to achieve a reduction of complexity at minimal cost of accuracy. The most common approach is to reduce the dimensionality of the physical model through reasonable simplifications and volume averaging. The Doyle-Fuller-Newman \cite{Doyle1993} p2D model is such a reduction in complexity. With additional assumptions, mostly valid for small currents or gradients, the Single-Particle-Model considering electrolyte effects (SPMe) \cite{Marquis2019} can be derived. The corresponding picture is that all particles in each electrode are equal, and hence only a single particle has to be solved. In this work, we use the SPMe description of a battery, as it achieves acceptable computation times while maintaining the general physicality of the battery description. Some of the limitations are discussed later. For the presented simulations, we use the PyBaMM framework \cite{Sulzer2021}.

The SPMe model relies on many parameters that describe the battery cell in general. As the cell was not explicitly characterized for the SPMe type of model, some parameters are unknown. The known parameters for this cell, e.g., OCV curves of silicon, graphite, and NCA, are extracted or calculated from the corresponding literature for this cell \cite{Wildfeuer2023, Schmitt2022, Kitada2019}. Remaining gaps are filled with literature values from cells that are very similar \cite{Chen2020, Ai2022, Kim2011}. The estimated mean cell temperatures (see Sec. \ref{subsec:ExpData}) are used to simulate the individual cell protocols. A table of the model parameters and their origin is given in the SI. However, slight errors in the parameter set are inevitable. Therefore, we adjusted some parameters slightly to match the voltage curves of the check-up cell as closely as possible (see SI for the parameter set). 
Further, we neglect the silicon voltage hysteresis \cite{Kbbing2024, Kbbing2024b} in our model for simplicity and because of its estimated low impact on the cell behavior with Si-Gr composite anodes, given the low silicon share.
Some alignment errors in the OCVs and overpotential deviations persisted, likely due to general model uncertainties.

\begin{table*}[]
\centering
\begin{tabular}{||c || c | c | c | c | c | c|| c | >{\centering\arraybackslash}p{1.1cm} | >{\centering\arraybackslash}p{1.1cm} | >{\centering\arraybackslash}p{1.4cm}| c ||} 
 \hline
 \multirow{2}{*}[-1.5pt]{Cell ID} &  \multirow{2}{*}[-1.5pt]{Chamber T.} & \multirow{2}{*}[-1.5pt]{Cell T.} & \multirow{2}{*}[-1.5pt]{Mean SoC} & \multirow{2}{*}[-1.5pt]{DoD} &  \multirow{2}{*}[-1.5pt]{\begin{tabular}{@{}c@{}}Charging \\ rate\end{tabular}} & \multirow{2}{*}[-1.5pt]{\begin{tabular}{@{}c@{}}Discharging \\ rate\end{tabular}} & \multirow{2}{*}[-1.5pt]{CL - $\textrm{LAM}_\textrm{Si}$} & \multicolumn{3}{c|}{CL - continuous SEI}  & \multirow{2}{*}[-1.5pt]{\begin{tabular}{@{}c@{}}CL - inst. \\ SEI\end{tabular}} \\ [0.1ex] \cline{9-11} &&&&&&&&total& regular& additional&\\ [0.1ex] 
 \hline
 \hline 
 249 & 20 °C & 26 °C& 40 \% & 80 \% & 0.5C & 1C & 7.3\% & 20.1\% & 12.8\% & 7.3\% & 72.6\% \\
 250 & 20 °C & 28 °C& 60 \% & 40 \% & 1C & 2C & 6.8\% & 48.6\% & 37.0\%& 11.6\% & 44.6\% \\
 288 & 20 °C & 27 °C& 40 \% & 5 \% & 1.5C & 1C & 6.9\% & 39.1\%& 29.9\%& 9.2\%& 54.0\%\\
 291 & 20 °C & 25 °C& 40 \% & 5 \% & 0.5C & 1C & 7.2\% & 33.1\%& 27.0\%& 6.1\% & 59.7\% \\
 389 & 20 °C & 24 °C& 50 \% & 100 \% & - & - & 7.5\% & 17.6\%& 14.0\%& 3.6\% & 74.9\% \\
 426 & 20 °C & 28 °C& 40 \% & 20 \% & 2C & 1C & 6.6\% & 38.7\%& 32.8\%& 5.9\% & 54.7\%\\
 429 & 20 °C & 28 °C& 50 \% & 100 \% & 1C & 1C & 7.6\% & 19.1\%& 14.2\%& 4.9\% & 73.3\%\\ [0.1ex] 
 \hline\hline
 217 & 35 °C & 44 °C& 50 \% & 60 \% & 2C & 2C & 9.7\% & 42.7\% &38.6\%& 4.1\% & 47.6\% \\
 248 & 35 °C & 41 °C& 45 \% & 20 \% & 1C & 1C & 6.2\% & 43.5\% &33.9\%& 9.6\% & 50.3\% \\
 403 & 35 °C & 45 °C& 40 \% & 20 \% & 2C & 1C & 5.9\% & 44.5\% &37.4\%& 7.1\% & 49.6\% \\
 428 & 35 °C & 41 °C& 45 \% & 90 \% & 1C & 1C & 7.2\% & 21.7\% &16.0\%& 5.7\% & 71.1\% \\ [0.1ex] 
 \hline\hline
 406 & 50 °C & 58 °C& 40 \% & 20 \% & 2C & 1C & 4.8\% & 53.5\% &42.7\%& 10.8\% & 41.7\% \\ [0.1ex] 
 \hline
\end{tabular}
\caption{The experimental cells with moderate conditions and their corresponding cycling protocols, to which the model is compared. The cell ID refers to the identifier used in the original publication of the data (see Tab. 1 in \cite{Karger2024}). Cell C389 is the CU cell where back-to-back CUs were performed. The corresponding protocol for cell C389 is described in Sec. \ref{subsec:ExpData}. The last three columns describe the different contributions to the CL ($\textrm{LAM}_\textrm{Si}$, continuous SEI growth, instantaneous SEI growth; at the last available data point) of each cell predicted by the model. The contribution of "continuous SEI" is further divided into SEI on regular and additional surface, including the additional active area of the particle cracking. In the SI, we analyze additional cells at more challenging conditions (e.g., high C-rates or temperatures).}
\label{tab.Cycl_Protocols}
\end{table*}

\subsection{Degradation Mechanisms}\label{subsec:Degradation mechanisms}

In this work, we consider the following physicochemical degradation mechanisms: continuous SEI growth and particle cracking, which eventually leads to $\textrm{LAM}_\textrm{Si}$, surface modifications, and instantaneous SEI growth after crack formation. These will be explained below. To avoid overfitting with too many possible degradation mechanisms, we neglect Li-plating and electrolyte dry-out, which we expect to have only minor influence (see \citet{Kupper2018} and the elucidation in Sec. \ref{sec:Results}). Nevertheless, we discuss possible consequences from that in Sec. \ref{sec:Results}.

\subsubsection{Continuous SEI growth}

When the electrolyte comes into contact with the negative electrode, electrons from the electrode react with electrolyte molecules, consuming Li-ions to form a SEI layer on the negative electrode \cite{Peled2017}. The SEI layer prevents further contact but continues to grow over time due to the transport of SEI educts through the SEI layer, thereby reducing the battery's capacity and power. Several publications \cite{Single2018, Köbbing2023, Kolzenberg2020, Li2015} have investigated different transport models that aim to explain long-term SEI growth during battery storage. Focusing on different aspects, we demonstrated that the electron diffusion model accurately describes SEI growth and captures the central dependence on the negative electrode voltage \cite{Single2018, Köbbing2023, Philipp2025}. Also, experimental results indicate that an electron-diffusive mechanism is responsible for the long-term SEI growth \cite{Krauss2024}. Usually, in an SPMe-type model, SEI growth is described by the current density of electrons that drive SEI-forming reactions. According to \citet{Kolzenberg2020}, the current density due to electron diffusion is given by 

\begin{equation}\label{eq.j_ED}
    j_\mathrm{ED} = \frac{c_\mathrm{e^-}D_\mathrm{e^-}F}{L_\mathrm{SEI}}\mathrm{exp}\left(-\tilde{\eta}_\mathrm{SEI}\right),
    \end{equation}
where $D_\mathrm{e^-}$ is the diffusion constant of electrons through the SEI layer, $c_\mathrm{e^-}$ is the reference electron concentration at an electrode potential of 0V vs Li metal, $F$ is the Faraday constant, and $L_\mathrm{SEI}$ is the thickness of the SEI. Moreover, $\tilde{\eta}_\mathrm{SEI}$ is the SEI overpotential, which affects the amount of electrons available in the SEI at the electrode-SEI interface. This is given by
\begin{equation}
    \tilde{\eta}_\mathrm{SEI} = \frac{F}{RT}\left(\eta_\mathrm{int} + U_0(\mathrm{SoC}) + \frac{\mu_\mathrm{Li,0}}{F}\right),
\end{equation}
where $R$ is the universal gas constant, $T$ is the temperature, $\mu_\mathrm{Li,0}$ is the reference chemical potential, $U_0(\mathrm{SoC})$ is the open-circuit voltage of the blend-anode, and $\eta_\mathrm{int}$ is the intercalation overpotential. Assuming standard Butler-Volmer kinetics, the intercalation current density $j_\mathrm{int}$ in relation to the exchange current density $j_0$ gives the intercalation overpotential
\begin{equation}
    \eta_\mathrm{int} = \frac{2RT}{F} \mathrm{arcsinh}\left(\frac{j_\mathrm{int}}{2j_0}\right).
\end{equation}
The thickness of the SEI is then evolving according to the current density via
\begin{equation}\label{eq.L_SEI}
    \frac{\mathrm{d}L_\mathrm{SEI}}{\mathrm{d}t} = \frac{V_\mathrm{SEI}}{zF}j_\mathrm{ED},
\end{equation}
where $V_\mathrm{SEI}$ is the mean molar volume of the SEI molecule and $z$ is the number of electrons involved in the SEI formation reaction. 

We model continuous SEI growth on the entire active surface area of the negative electrode (see Fig. \ref{fig.Crack-Illustration}a)), as described by equations (\ref{eq.j_ED}) and (\ref{eq.L_SEI}). Silicon and graphite have different electrochemical properties and, therefore, presumably also varying SEI composition. Still, to avoid overfitting and lack of individual markers, we assume the SEI on graphite and silicon to be similar.

\subsubsection{Particle cracking}\label{subsec:Part.Cracking}

Due to the significant volume expansion during lithiation and delithiation \cite{Boukamp1981}, silicon particles experience substantial mechanical stresses. Periodic cycling means repeated mechanical loads, which cause material fatigue and, consequently, cracks.

In materials science, material fatigue is statistically analyzed using Wöhler curves \cite{Basquin1910}, which plot the number of cycles to failure as a function of stress amplitude. The repeated application of stress amplitudes parallels the continuous cycling of battery materials, and hence, the crack opening and growth is often related to particles' tensile stresses in the literature \cite{Laresgoiti2015, Deshpande2012, Reniers2019} (see Fig. \ref{fig.Crack-Illustration}a)). \citet{Bonkile2024} used the stress amplitude during cycles to describe the LAM due to cracking. Because the underlying physical mechanisms for LAM and cracking are similar \cite{Reniers2019}, we adopt this equation to describe crack evolution. In the SPMe picture of a battery, a single particle crack can be defined with a fixed crack width $b_{cr}$ and a variable crack length $l_\mathrm{cr}$ (see Fig. \ref{fig.Crack-Illustration}d)). Then, the evolution of a crack is given by
\begin{equation}\label{eq.crack-evo}
    \begin{split}
    \partial_t l_\mathrm{cr} &= \beta \left(\frac{\sigma_\mathrm{h,surf}-\sigma_\mathrm{h,surf,min}}{\sigma_\mathrm{crit}}\right)^{m} \\
    &= k_\mathrm{cr} \sigma_\mathrm{t,surf}^{m},
    \end{split}
\end{equation}
with the critical stress $\sigma_\mathrm{crit}$ and the hydrostatic stress $\sigma_\mathrm{h,surf}$ at the surface, which is given by
\begin{equation}
    \sigma_\mathrm{h,surf} = \frac{2 \sigma_\mathrm{t,surf} + \sigma_\mathrm{r,surf}}{3}.
\end{equation}
It depends on tangential ($\sigma_\mathrm{t,surf}$) and radial ($\sigma_\mathrm{r,surf}$ = 0) stresses at the surface of the particle, which were derived for spherical particles \cite{Dai2014, Wu2017, Bohn2013, Li2017, Fu2013}. The minimal hydrostatic stress is assumed to be zero for complete cycles ($\sigma_\mathrm{h,surf,min} = 0$). Inserting this into the first line of eq. (\ref{eq.crack-evo}) yields the second line, with a rate constant $k_\mathrm{cr}$ and the stress exponent $m$. The tangential stress at the surface is given by

\begin{equation}\label{eq.tang-stress}
    \sigma_\mathrm{t,surf} = \frac{V_\textrm{Li} E}{3(1-\nu)} (\overline{c_\mathrm{s}} - c_\mathrm{s,surf}),
\end{equation}
where $V_\textrm{Li}$ is the partial molar volume of lithium, $E$ is the Young's modulus, $\nu$ is the Poisson's ratio, $\overline{c_\mathrm{s}}$ is the particle spatial averaged concentration, and $c_\mathrm{s,surf}$ is the concentration at the surface of the particle.

\begin{figure*}[]
	\centering
	\makebox[0pt]{%
    \includegraphics[width=0.85\paperwidth]{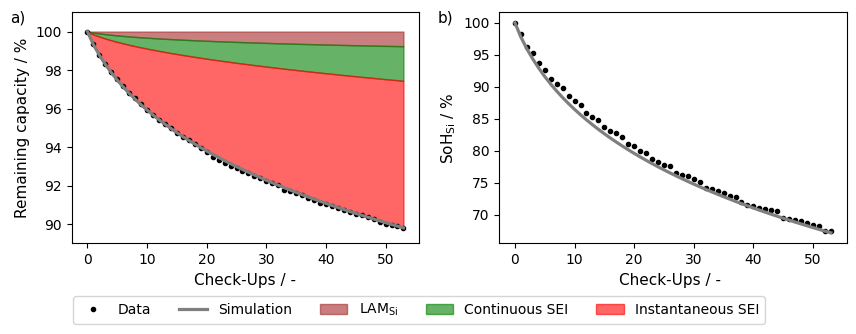}}
	\caption{a) CL and b) $\mathrm{SoH}_\mathrm{Si}$ of the continuous check-up cycling cell C389. The colored areas in a) describe the different contributions to the simulated CL according to the model.}
	\label{fig.C389-CL-SoH}
\end{figure*}

The reported cracking behavior of silicon in the literature ranges from slow dissolution and SEI growth into the particles \cite{He2021} to complete particle pulverization \cite{Wetjen2018} (see Fig. \ref{fig.Crack-Illustration}b) and c)). This crack model can describe both: many nano-cracks at the silicon particle surface slowly progressing into the particle core, as well as large particle cracks. Our model does not distinguish between these effects. We model the occurrence of particle cracks to have three different consequences: $\textrm{LAM}_\textrm{Si}$, particle surface modifications, and the initiation of instantaneous SEI growth. We discuss them in the following.

\textbf{$\textrm{LAM}_\textrm{Si}$}: Progressive crack growth can lead to island formation (see Fig. \ref{fig.Crack-Illustration}b)) and, consequently, to the loss of electrical contact to parts of the silicon particles (see Fig. \ref{fig.Crack-Illustration}b-c)). We model the $\textrm{LAM}_\textrm{Si}$ through the evolution of the corresponding active material volume fraction $\epsilon_\mathrm{Si}$ as
\begin{equation}\label{eq.LAM}
    \partial_t \epsilon_\mathrm{Si} = - k_\mathrm{LAM} \sigma_\mathrm{t,surf}^{m}\epsilon_\mathrm{Si},
\end{equation}
with a rate constant $k_\mathrm{LAM}$. As we assume that cracking and $\textrm{LAM}_\textrm{Si}$ rely on the same physical mechanism, we also assume the exponent $m$ to be identical for both processes, in contrast to the literature \cite{Reniers2019}. Then, $\textrm{LAM}_\textrm{Si}$ can be expressed by
\begin{equation}
    \epsilon_\mathrm{Si} = \epsilon_\mathrm{Si,0} \mathrm{exp}\left(-\frac{ k_\mathrm{LAM}}{k_\mathrm{cr}} l_\mathrm{cr}\right),
\end{equation}
with the initial active material fraction of silicon $\epsilon_\mathrm{Si,0}$. Previously (see eq.(\ref{eq.crack-evo})), we discussed the growth of a single crack only. Here in eq. (\ref{eq.LAM}), we describe the contribution of all occurring cracks on all particles to the total $\textrm{LAM}_\textrm{Si}$. The total number of cracks is proportional to the active surface area and, therefore, the available silicon material. Hence, we model $\textrm{LAM}_\textrm{Si}$ as proportional to the amount of available active material, $\epsilon_\mathrm{Si}$. Assuming cracks grow from the outside inwards, the remaining active particle size should decrease. However, our model does not account for changes in particle size.

\textbf{Particle surface modifications}: We model that particle cracks generate two distinct additional surface areas of the silicon particles. First, we consider a fresh electrode area $A_\mathrm{Si,crack}$ inside the particles, which does not contribute to the total active area but is directly exposed to electrolyte (see Fig. \ref{fig.Crack-Illustration}d)). Therefore, $A_\mathrm{Si,crack}$ will be covered by instantaneous SEI growth. Second, an additional active area of the silicon particles $A_\mathrm{Si,add.}$ is relevant for exchange currents (see Fig. \ref{fig.Crack-Illustration}d)) and continuous SEI growth.

The area of two-sided cracks inside the particles is given by $2l_\mathrm{cr}b_\mathrm{cr}n_\mathrm{cr}$ with the variable crack length $l_\mathrm{cr}$, the fixed crack width $b_\mathrm{cr}$ and number of cracks $n_\mathrm{cr}$. Assuming that cracks are evenly distributed over the basic silicon area, the number of cracks follows $n_\mathrm{cr} = \rho_\mathrm{cr} A_\mathrm{Si,basic}$, with the area specific crack number density $\rho_\mathrm{cr}$ and basic silicon area $A_\mathrm{Si,basic}$. The basic surface area of the silicon particles can be calculated via the surface area per volume density $a_\mathrm{Si}$ and volume of the negative electrode $V_\mathrm{NE}$ so that $A_\mathrm{Si,basic} =  a_\mathrm{Si} V_\mathrm{NE}$. For spherical silicon particles with radius $r_\mathrm{Si}$, the surface area per volume density is $a_\mathrm{Si} = \frac{3 \epsilon_\mathrm{Si}}{r_\mathrm{Si}}$. Then, the total cracked area inside the particle is given by
\begin{equation}\label{eq.A_Si_crack}
    A_\mathrm{Si,crack} = \frac{3 \epsilon_\mathrm{Si}}{r_\mathrm{Si}}2 l_\mathrm{cr} b_\mathrm{cr} \rho_\mathrm{cr}V_\mathrm{NE}.
\end{equation}
We model that thin cracks in the particle with a certain opening angle increase the total active area only by a small fraction $f$ of the crack area inside the particle, i.e., $A_\mathrm{Si,add.} = \frac{A_\mathrm{Si,crack}}{f}$. This concept can be concluded in the surface roughness of silicon $\xi_\mathrm{Si}$, which is given by 
\begin{equation}
    \xi_\mathrm{Si} = 1 + \frac{2 l_\mathrm{cr} b_\mathrm{cr} \rho_\mathrm{cr}}{f},
\end{equation}
and modifies the total active surface area of silicon through
\begin{equation}\label{eq.A_Si_tot}
    \begin{split}
        A_\mathrm{Si,total} & = \xi_\mathrm{Si}A_\mathrm{Si,basic} \\
         & = \frac{3 \epsilon_\mathrm{Si}}{r_\mathrm{Si}}V_\mathrm{NE} + \frac{3 \epsilon_\mathrm{Si}}{r_\mathrm{Si}}\frac{2 l_\mathrm{cr} b_\mathrm{cr} \rho_\mathrm{cr}}{f}V_\mathrm{NE} \\
         & = A_\mathrm{Si,basic} + A_\mathrm{Si,add.}.
    \end{split}
\end{equation}
So, the crack growth increases the additional active area of a silicon particle and the area exposed to the electrolyte.

\textbf{Instantaneous SEI growth}: On the fresh electrode area exposed to the electrolyte $A_\mathrm{Si,crack}$, SEI will grow very fast, as there is no passivation (see Fig. \ref{fig.Crack-Illustration}c-d)). We model it as growing instantaneously to a specific thickness $d_\mathrm{SEI}$, which is only valid for a low SEI thickness (i.e., a few nanometers). Further, we assume that the cracks within the particle provide limited space and electrolyte volume for SEI growth. This means that we model the growth of the SEI on the crack area inside the particles as an instantaneous process, with vanishing ongoing SEI growth due to the restricted volume within the cracks. Thereby, the crack growth is accompanied by direct CL due to SEI formation. This CL is related to the volume of the SEI molecules formed on cracks $V_\mathrm{SEI,cracks}$. With $I_\mathrm{crack} = \partial_t Q_\mathrm{crack} = \frac{F}{V_\mathrm{SEI}}\partial_t V_\mathrm{SEI,cracks}$ the corresponding current density for SEI growth on cracks is
\begin{equation}\label{eq.j_SEICrack}
\begin{split}
        j_\mathrm{SEI,cracks} &= \frac{I_\mathrm{crack}}{A_\mathrm{Si,crack}} = \frac{\frac{F}{V_\mathrm{SEI}} \partial_t (A_\mathrm{Si,crack}d_\mathrm{SEI})}{A_\mathrm{Si,crack}} \\
        & \approx \frac{F}{V_\mathrm{SEI}} \frac{\partial_t l_\mathrm{cr}}{l_\mathrm{cr}} d_\mathrm{SEI}.
\end{split}
\end{equation}
Details regarding the derivation of the second line are shown in the SI. Equation (\ref{eq.j_SEICrack}) describes the electron current density to form a SEI layer with fixed thickness $d_\mathrm{SEI}$ while the crack is growing. 

In contrast, on the additional active surface area $A_\mathrm{Si,add.}$ (see Fig. \ref{fig.Crack-Illustration}d)), the SEI growth is modeled according to eq. (\ref{eq.j_ED}), for which we estimate conservatively the thickness of the SEI to be identical to the SEI thickness on the basic surface $A_\mathrm{Si,basic}$. We investigate the model inaccuracy introduced by this assumption in the SI.

To conclude, we model the particle cracking to have three distinct effects. First, the $\textrm{LAM}_\textrm{Si}$, which is given by eq. (\ref{eq.LAM}). Second, particle surface modification, which describes on the one hand the additional active area of silicon particles available for lithium insertion (see eq. (\ref{eq.A_Si_tot})) and continuous SEI growth (described by eq. (\ref{eq.j_ED})), and on the other hand the area exposed to electrolyte in the particle cracks (see eq. (\ref{eq.A_Si_crack})). Third, the instantaneous SEI growth inside the particle cracks, which is described by eq. (\ref{eq.j_SEICrack}).

\subsubsection{Parameters}

Given the importance of avoiding overfitting, we summarize here the parameters required for the degradation model and how they are estimated. The model uses a single parameter set to describe the experimental data across all different usage conditions.

\begin{itemize}
    \item continuous SEI growth: determined by the diffusion coefficient of electrons $D_\mathrm{e^-}$ in eq. (\ref{eq.j_ED}), which is adapted through an Arrhenius dependency with an activation energy $E_A$ for cells at higher temperatures: $D_\mathrm{e^-} = D_\mathrm{e^-,ref.}\mathrm{exp}(\frac{E_A}{R}(\frac{1}{T_\mathrm{ref}}-\frac{1}{T}))$.
    \item particle cracking: given by the crack rate constant $k_\mathrm{cr}$ and the silicon stress exponent $m$ in eq. (\ref{eq.crack-evo}).
    \item $\textrm{LAM}_\textrm{Si}$: described by the LAM rate constant $k_\mathrm{LAM}$ and again the silicon stress exponent $m$ in eq. (\ref{eq.LAM}).
    \item instantaneous SEI growth: depends on the thickness of the instantaneously grown SEI after crack formation $d_\mathrm{SEI}$ in eq. (\ref{eq.j_SEICrack}) and the exposed crack area due to particle cracking.
\end{itemize}

We extract the reference diffusivity of electrons through the SEI $D_\mathrm{e^-,ref.}$ from the capacity loss measured in a storage experiment without intermediate CUs at $T = 20$°C. Then, we obtain the activation energy $E_A$ from storage cells at elevated temperatures (see SI). The remaining parameters are adjusted to meet the behavior of the cells stated in Tab. \ref{tab.Cycl_Protocols} as described in the following. The cracking rate $k_\mathrm{cr}$ and the inherent silicon mechanical fatigue parameter $m$ are fitted so that the increased active area captures the decrease in stress and the mitigation behavior of $\textrm{LAM}_\textrm{Si}$ during battery cycling. The LAM rate $k_\mathrm{LAM}$ quantifies the degree to which cracks that occur lead to $\textrm{LAM}_\textrm{Si}$ and is fitted to observations of $\mathrm{SoH}_\mathrm{Si}$ during cycling. A certain amount of CL by instantaneous SEI growth is determined by the number of formed SEI molecules $n_\mathrm{SEI,cracks}$ and hence related to the mean molar volume of SEI molecules $V_\mathrm{SEI}$, the thickness of the instantaneously grown SEI after crack formation $d_\mathrm{SEI}$, and the actual area exposed to the electrolyte. This area is determined by the number of cracks and the increasing area exposed to electrolyte per crack (through the crack width $b_\mathrm{cr}$, and the increasing crack length $l_\mathrm{cr}$). In a static picture (without LAM), the CL by instantaneous SEI growth is then described by
\begin{equation}
        n_\mathrm{SEI,cracks} = \frac{V_\mathrm{SEI,cracks}}{V_\mathrm{SEI}} = \frac{2 l_\mathrm{cr} b_\mathrm{cr} \rho_\mathrm{cr} d_\mathrm{SEI}}{V_\mathrm{SEI}}\frac{3 \epsilon}{r_\mathrm{Si}}V_\mathrm{NE}.
\end{equation}
Since all factors depend on each other multiplicatively, the CL by instantaneous SEI growth is effectively described by only one parameter. Therefore, we fixed all constants except $d_\mathrm{SEI}$, which is fitted to the remaining difference in observed CL and simulated CL by $\textrm{LAM}_\textrm{Si}$ and continuous SEI growth. 

To sum up, five parameters are needed in the degradation model: the reference diffusion coefficient of electrons $D_\mathrm{e^-,ref.}$, the crack rate constant $k_\mathrm{cr}$, the silicon stress exponent $m$, the LAM rate constant $k_\mathrm{LAM}$, and the thickness of the instantaneously grown SEI after crack formation $d_\mathrm{SEI}$. Extending the degradation model to higher temperatures, the activation energy $E_A$ is needed as sixth parameter.

\section{Results \& Discussion}\label{sec:Results}

To understand the degradation observed in the storage experiments, it is essential to capture the degradation caused by battery cycling during the CUs. Therefore, we first examine the cycling results and then investigate the storage results.

\subsection{Cycling}

The results for the cells cycled at moderate temperatures ($T = 20$°C and $T = 35$°C)  and C-rates ($\leq 2$C) are summarized in Fig. \ref{fig.Cycl-Summary-CL-SoH}. It includes 11 cells in total, with the individual protocols listed in Tab. \ref{tab.Cycl_Protocols}. The figure reveals that the model captures the CL and $\mathrm{SoH}_\mathrm{Si}$ well across various cycling conditions, with overall low RMSE (1.11\% in CL and 2.50\% in $\textrm{SoH}_\textrm{Si}$).
In the following, we investigate individual cells to show the different contributions to the CL predicted by the model and their dependence on operating conditions. We also examine the model's limitations for some cycling protocols.

\begin{figure}[]
	\centering
	\makebox[0pt]{%
    \includegraphics[width=0.41\paperwidth]{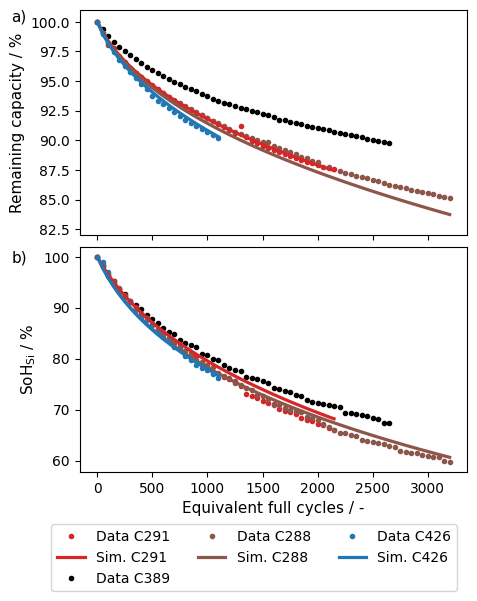}}
	\caption{a) CL and b) $\mathrm{SoH}_\mathrm{Si}$ of cells cycled in the intermediate SoC region with $T = 20$°C (C288, C291, C426). The data of the CU cell C389 (black dots) is included to visualize the amount of degradation solely due to the number of performed CUs. Note that these do not refer to the cycled EFCs but correspond to the number of CUs of the cycling cells.}
	\label{fig.Summary-MediumSoCT20}
\end{figure}

\begin{figure*}
	\centering
	\makebox[0pt]{%
    \includegraphics[width=0.85\paperwidth]{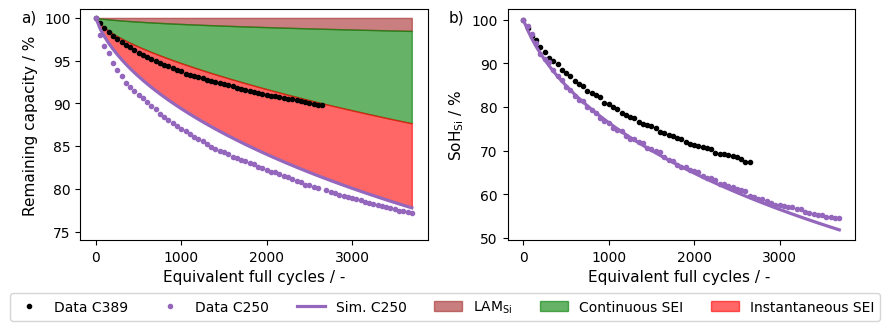}}
	\caption{a) CL and b) $\mathrm{SoH}_\mathrm{Si}$ of cell C250, cycled at elevated SoC. The colored areas in a) describe the different contributions to the simulated CL described by the model. The data of the CU cell C389 (black dots) is included to visualize the amount of degradation solely due to the number of performed CUs. Note that these do not refer to the cycled EFCs but correspond to the number of CUs of the cycling cells.}
	\label{fig.Summary-CompC250}
\end{figure*}

\subsubsection{Degradation during check-ups}

As each cell undergoes several CUs (every 50 EFCs), they can significantly contribute to cell degradation and affect the degradation pathways. To disentangle the degradation measured during regular cycling from that solely due to the CUs, we first investigate the CUs' impact.

Figure \ref{fig.C389-CL-SoH} shows the measured and simulated CL and $\mathrm{SoH}_\mathrm{Si}$ for the cell undergoing 54 subsequent CUs (cell C389), without intermediate cycling between the CUs. Our model captures the trends in CL and $\mathrm{SoH}_\mathrm{Si}$ very well. In Fig. \ref{fig.C389-CL-SoH}a), the simulated contributions to CL ($\textrm{LAM}_\textrm{Si}$, continuous SEI growth, and instantaneous SEI growth) are shown. The continuous SEI growth on the additional active area due to particle cracking is included in the continuous SEI growth. Here, the dominant share is the degradation due to instantaneous SEI growth after crack formation in the particles. The degradation due to continuous SEI growth during the 54 CUs, which is similar to $\sim$160 equivalent full cycles (EFCs), accounts for only 1-2\% of the lost capacity. The share of CL due to $\textrm{LAM}_\textrm{Si}$ is even smaller.

Figure \ref{fig.C389-CL-SoH}b) depicts the remaining active silicon $\textrm{SoH}_\textrm{Si} = 1 - \textrm{LAM}_\textrm{Si}$ depending on the number of CUs. Experiment and simulation show severe decrease of active silicon, reaching $\textrm{SoH}_\textrm{Si} = 70\%$ after 50 CUs. The $\textrm{LAM}_\textrm{Si}$ itself does not contribute directly to the CL, as LLI is the dominant cause for CL. However, the lithium lost together with the silicon contributes to the LLI and CL, depending on the degree of lithiation. As shown in Fig. \ref{fig.C389-CL-SoH}a) this contribution is rather small. Note, even losing silicon ($\textrm{LAM}_\textrm{Si}$) in the fully lithiated state can not alone explain the large CL due to the low total silicon capacity ($\approx$11\%).

Note the very similar trend in the total CL and $\mathrm{SoH}_\mathrm{Si}$, which indicates that they are mainly driven by the same mechanism, namely particle cracking. In our model, the profile of $\mathrm{SoH}_\mathrm{Si}$ contains the information about the amount of mechanical damage. Why is the profile of $\mathrm{SoH}_\mathrm{Si}$ flattening? The cracking leads, on the one hand, to additional surface area with significant CL due to instantaneous SEI growth, and, on the other hand, to loss of electrical contact to parts of the silicon particles. The predominant increase in active area reduces the interfacial current density because the current is distributed over a larger area. In the subsequent discharge, concentration gradients and the stresses are therefore lower, leading to the observed self-mitigating effect in the model. In addition, the shrinkage of the active silicon particle by $\textrm{LAM}_\textrm{Si}$ of the outer parts would also lead to decreasing concentration gradients due to faster diffusion time ($\tau = \frac{r_\mathrm{Si}^2}{D_\mathrm{Si}}$). However, in the model, we assume a fixed particle size to reduce complexity.

The crack opening or its evolution is induced by tensile tangential stress occurring in silicon particles due to concentration gradients during discharge \cite{Liu2012}. The buildup of concentration gradients in the silicon particles depends on the actual use of silicon during charging or discharging. In a silicon-graphite composite electrode, the alignment of the OCV curves of the individual materials, combined with the emerging overpotentials during use, determines which material is preferably used. For silicon, this is the case in the lower SoC region (e.g., see alignment of the qOCVs in \citet{Wildfeuer2022}), so silicon is repeatedly used during the CU cycling. In the following, we investigate the degradation of cells cycled only in the intermediate SoC region with substantially less usage of silicon.

\subsubsection{Cycling in intermediate SoC regions}\label{subsec:Interm-Cycling}

\begin{figure}[]
	\centering
	\makebox[0pt]{%
    \includegraphics[width=0.41\paperwidth]{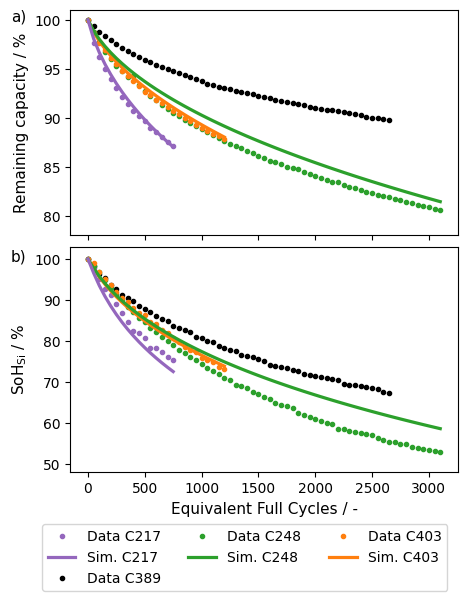}}
	\caption{a) CL and b) $\mathrm{SoH}_\mathrm{Si}$ of cells cycled in the intermediate SoC region with $T = 35$°C (C217, C248, C403). The data of the CU cell C389 (black dots) is included to visualize the amount of degradation solely due to the number of performed CUs. Note that these do not refer to the cycled EFCs but correspond to the number of CUs of the cycling cells.}
	\label{fig.Summary-MediumSoCT35}
\end{figure}

\begin{figure}[]
	\centering
	\makebox[0pt]{%
    \includegraphics[width=0.41\paperwidth]{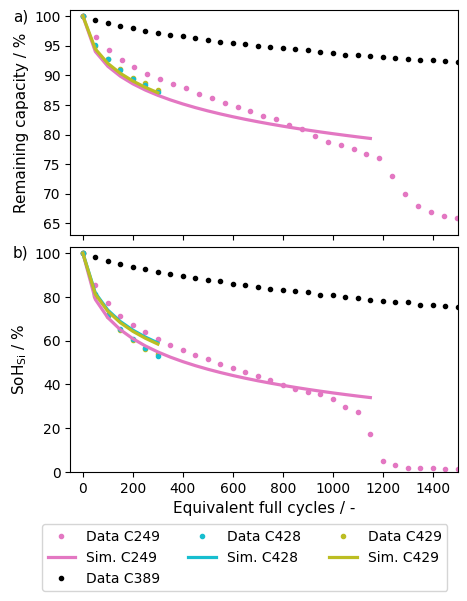}}
	\caption{a) CL and b) $\mathrm{SoH}_\mathrm{Si}$ of cells cycled with high DoD $\geq 80$\% (C249, C428, C429). The data of the CU cell C389 (black dots) is included to visualize the amount of degradation solely due to the number of performed CUs. Note that these do not refer to the cycled EFCs but correspond to the number of CUs of the cycling cells.}
	\label{fig.Summary-HighDoD}
\end{figure}

\begin{figure}[]
	\centering
	\makebox[0pt]{%
    \includegraphics[width=0.41\paperwidth]{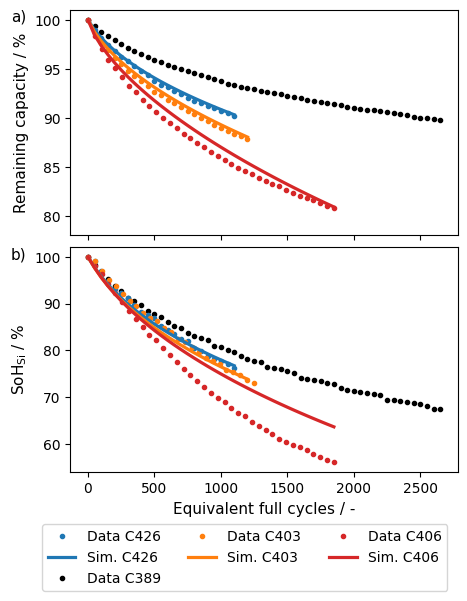}}
	\caption{a) CL and b) $\mathrm{SoH}_\mathrm{Si}$ of cells with the identical cycling protocol in the intermediate SoC region but varying temperatures $T = 20$°C (C426), $T = 35$°C (C403), and $T = 50$°C (C406). The data of the CU cell C389 (black dots) is included to visualize the amount of degradation solely due to the number of performed CUs. Note that these do not refer to the cycled EFCs but correspond to the number of CUs of the cycling cells.}
	\label{fig.Summary-VaryingT}
\end{figure}

The results for the cells cycled in the intermediate SoC region at $T = 20$°C (cells C288, C291, C426) are shown in Fig. \ref{fig.Summary-MediumSoCT20}. The black dots represent the data from the CU cell C389 and are included to visualize the degradation that would occur solely due to the CU. Therefore, the black dots do not refer to the number of EFCs, but to the number of performed CUs (one CU every 50 EFCs). The CL is slightly increased, while the $\mathrm{SoH}_\mathrm{Si}$ remains very close to that of the CU only cell (see black dots), despite the cycled cells undergoing nearly 17 times more EFCs compared to the CU cell. The variation in the charging rate indicates that higher currents lead to slightly accelerated degradation. The model accurately captures both the trends and amplitudes of CL and $\mathrm{SoH}_\mathrm{Si}$ across all cycling protocols. Only for cell C288 does the model slightly overestimate the CL at high cycle numbers.

\begin{figure}[]
	\centering
	\makebox[0pt]{%
    \includegraphics[width=0.41\paperwidth]{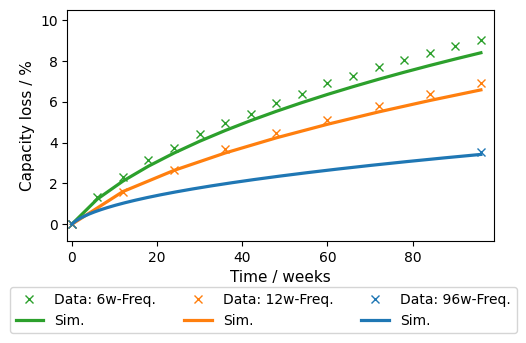}}
	\caption{CL of cells stored at 50\% SoC for 96 weeks at $T = 20$°C with varying CU frequency (every 6/12/96 weeks). The data point for storage with only one CU (blue cross) is used to identify the reference diffusion coefficient of electrons $D_\mathrm{e^-,ref.}$ for the continuous SEI growth (see eq. (\ref{eq.j_ED})).}
	\label{fig.Storage-SoC50-F6-12-96}
\end{figure}

The increased CL relative to the CU cell is captured in our model by continuous SEI growth during the numerous cycles between CUs. In contrast, the limited silicon utilization in the intermediate SoC range results in negligible crack formation, leading to minimal associated $\textrm{LAM}_\textrm{Si}$ and instantaneous SEI formation. Consequently, the main degradation in these cells is crack formation during the CUs and continuous SEI growth during cycling in the intermediate SoC regions. The electron diffusion mechanism describing continuous SEI growth successfully captures the influence of the C-rate on the rate of SEI growth via eq. \ref{eq.j_ED} and serves as a valid SEI model parameterized by a single storage experiment only (see Fig. \ref{fig.Storage-SoC50-F6-12-96} in Sec. \ref{subsec:storage}). One possible reason for the slightly overestimated CL in cell C288 is an overestimation of the additional active area to grow continuous SEI. We will discuss the additional area in Sec. \ref{subsec:storage}, Sec. \ref{subsec:limits}, and in the SI Sec. VI.

To further demonstrate that continuous SEI growth is the dominating degradation mechanism during regular cycling, we decompose the individual contributions to the CL for cell C250, which is cycled in a slightly elevated SoC regime. The results are shown in Fig. \ref{fig.Summary-CompC250}. The comparison shows a slight underestimation of the CL, but overall good agreement between the simulated CL and $\mathrm{SoH}_\mathrm{Si}$ and the experimental data. The contribution of continuous SEI growth to the CL is substantially larger than in the CU cell C389 (see Fig. \ref{fig.C389-CL-SoH}a)). Notably, the absolute contribution of instantaneous SEI growth to CL remains comparable to that in cell C389 (see red areas in Fig. \ref{fig.C389-CL-SoH}a) and Fig. \ref{fig.Summary-CompC250}b)). This reaffirms that for cells cycled in the intermediate SoC region, degradation is primarily governed by the continuous SEI growth during regular cycling and crack formation during CU cycles. Given the model’s strong performance in capturing degradation in this regime, we proceed to investigate the influence of slightly elevated temperatures in the following.

Results for cells cycled in the intermediate SoC region at elevated temperature ($T = 35$°C, cells C217, C248, C403) are shown in Fig. \ref{fig.Summary-MediumSoCT35}. For cells C217 and C403, the simulated CL and $\mathrm{SoH}_\mathrm{Si}$ show excellent agreement with the experimental observations. The model slightly underestimates $\textrm{LAM}_\textrm{Si}$ and CL for cell C248, though the overall trends are well captured. Notably, cell C217 exhibits a significantly higher CL compared to the other cells, despite having a similar mean SoC. This enhanced degradation is attributed to a higher DoD and elevated charge and discharge rates. The increased DoD and charge current lead to more cycling in the higher SoC region with larger overpotentials resulting in increased continuous SEI growth (see eq. (\ref{eq.j_ED})). Additionally, the high DoD results in greater silicon utilization at low SoC, resulting in higher CL due to instantaneous SEI growth. The model successfully captures this behavior. To further validate the detrimental impact of high DoD (especially deep discharge into the low SoC regime) on cell degradation, we analyze cells subjected to high DoD cycling in the following section.

\subsubsection{Cycling with high DoD}\label{subsec:HighDoD}

The results for the cells cycled with high DoD ($\geq 80$\%) at $T = 20$°C (cells C249, C429) and $T = 35$°C (C428) are presented in Fig. \ref{fig.Summary-HighDoD}. Notably, the model accurately predicts the CL and $\mathrm{SoH}_\mathrm{Si}$ for cells C428 and C429, even under severe degradation conditions. For cell C249, the simulation captures the overall degradation trends well, though it slightly overestimates the CL during the first 1000 EFCs. Beyond this point, the cell exhibits irregular degradation behavior, which is not accounted for in the model and thus excluded from the model quality assessment (see Fig. \ref{fig.Cycl-Summary-CL-SoH}).

Similar to the CU cycling, the cells with high DoD become repeatedly cycled into the lower SoC region, where silicon is preferably used during discharge. This results in significant tensile stresses and crack formation within the silicon particles. The model's representation of particle cracking successfully explains the pronounced $\textrm{LAM}_\textrm{Si}$ and elevated CL observed in high-DoD cycled cells. In cell C249, additional non-homogeneous effects may contribute to the early onset of a knee-point, potentially accounting for the minor deviation in the initial degradation phase. 

A direct comparison between cell C250, cycled with 40\% DoD (see Fig. \ref{fig.Summary-CompC250}), and cell C249, cycled with 80\% DoD (see Fig. \ref{fig.Summary-HighDoD}), reveals that C250 exhibits approximately half the CL after 1000 EFCs, despite being subjected to twice the C-rate and cycled to the same high SoC limit. As cell C250 faces a higher risk of lithium plating but shows less CL, we conclude that plating's contribution to CL is negligible. This observation confirms that the severe degradation observed in high DoD cycling is due to the cycling into the lower SoC region, in which silicon is preferably used. It aligns with the well-established finding that using a higher lower-voltage cut-off, i.e. avoiding the low SoC range, can substantially improve the cycle life of silicon-containing anodes \cite{Zhang2011}.

\subsubsection{Cycling at elevated temperatures}

For further validation and to assess the model's limitations, we also examine cells cycled at high temperatures ($T \geq 50$°C). These cases are excluded from the model quality overview in Fig. \ref{fig.Cycl-Summary-CL-SoH} for two main reasons. First, higher temperatures can trigger multiple effects that are not explicitly accounted for in the current model. Second, model parameters were calibrated using data from cells tested at $T = 20$°C, and the temperature dependence of various battery processes (e.g., diffusion, reaction kinetics) is not explicitly characterized. Without accurate temperature dependencies, extrapolating the model to higher temperatures may introduce significant errors. Therefore, we first focus on a controlled comparison of cells subjected to the identical cycling protocol at different temperatures. Further cells cycled at high temperatures are analyzed in the SI Sec. V. 

Figure \ref{fig.Summary-VaryingT} compares the degradation behavior of cells cycled under identical protocols but at varying temperatures. The results show a clear trend: increasing temperature leads to increased CL and $\mathrm{LAM}_\mathrm{Si}$. The model captures the overall trend and magnitude of CL accurately across all temperatures. For $T = 20$°C and $T = 35$°C, the simulated $\mathrm{SoH}_\mathrm{Si}$ aligns perfectly with the experimental data. However, at $T = 50$°C, the model underestimates $\textrm{LAM}_\textrm{Si}$. This discrepancy may stem from inaccuracies in the assumed thermal dependencies of various cell processes, or temperature-dependent changes in the mechanical behavior of silicon. Since the CL fits quite well, it is also possible that an additional $\textrm{LAM}_\textrm{Si}$ effect occurs at higher temperatures, unrelated to cracking and accompanying SEI growth. Additionally, high temperatures might influence the amount of $\textrm{LAM}_\textrm{Si}$ due to cracks.

To conclude the investigation of cyclic aging, our proposed model successfully captures the strong degradation during CU cycling and cycling with high DoD, as well as the relatively mild degradation observed in the intermediate SoC range or at slightly elevated temperatures.

\begin{figure*}[]
	\centering
	\makebox[0pt]{%
    \includegraphics[width=0.85\paperwidth]{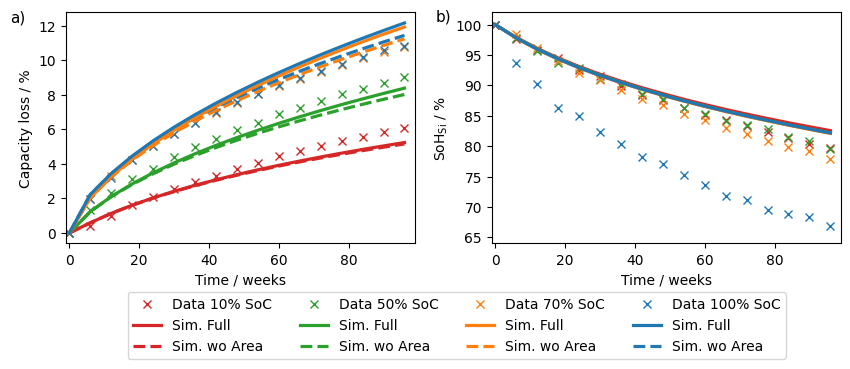}}
	\caption{a) CL and b) $\mathrm{SoH}_\mathrm{Si}$ of cells stored for 96 weeks at $T = 20$°C with different SoCs (10\%, 50\%, 70\%, and 100\%) and CUs every 6 weeks.}
	\label{fig.Storage-SoC-F6w}
\end{figure*}

\subsection{Storage}\label{subsec:storage}

Previously, we have demonstrated a strong agreement between model predictions and experimental cycling data. Now, we transfer this to storage data and investigate the influence of frequent CU cycling on the observed storage degradation.

Figure \ref{fig.Storage-SoC50-F6-12-96} presents the results of storage experiments conducted at 50\% SoC and $T = 20$°C, with varying CU-frequencies. The figure reveals that more CUs lead to more CL. Comparing the 6-week and 96-week frequency data reveals that degradation occurring during the CUs can dominate degradation during the storage periods. The model accurately reproduces the experimental trends across all CU frequencies. Notably, the observed reduction in degradation per individual CU (as shown in Fig. \ref{fig.C389-CL-SoH}) means that doubling the CU frequency (from 12-week to 6-week intervals) does not result in doubling the additional degradation. This nonlinear behavior is well captured by our model. For the highest CU-frequency, the model slightly underestimates the CL after the first year of storage.

For model parameterization, the data point corresponding to a single CU at the end of the 96-week storage period (blue cross in Fig. \ref{fig.Storage-SoC50-F6-12-96}) is used to determine the reference diffusion coefficient of electrons $D_\mathrm{e^-,ref.}$, which describes the continuous SEI growth. In Sec. V in the SI, we further demonstrate that the model successfully predicts storage behavior at $T = 35$°C across different CU-frequencies. The data from the single CU experiment at $T = 35$°C is used to determine the Arrhenius-dependency of $D_\mathrm{e^-}$.

The degradation behavior during the 96-week storage experiment with constant CU frequency and varying SoCs is summarized in Fig. \ref{fig.Storage-SoC-F6w}. The results show that higher SoC values lead to increased CL and $\textrm{LAM}_\textrm{Si}$. As shown in Fig. \ref{fig.Storage-SoC-F6w}a), the model accurately captures the experimental CL within the first year of storage across all SoCs. Beyond this, small deviations emerge: the model slightly overestimates CL at high SoCs (70\% and 100\%), while underestimating it at low SoCs (10\% and 50\%). Simulations that neglect SEI growth on newly exposed surface areas due to cracking (dashed lines) perform better at 70\% and 100\% SoC but worsen the underestimation at 10\% and 50\% SoC. Figure \ref{fig.Storage-SoC-F6w}b) illustrates a very similar simulated $\mathrm{SoH}_\mathrm{Si}$ across all SoC levels. The model predictions align well with experimental data for 10\%, 50\%, and 70\% SoC during the first year. However, after one year, $\mathrm{SoH}_\mathrm{Si}$ is slightly overestimated in all cases. At 100\% SoC, the deviation is particularly pronounced, with the simulation showing a clear misalignment with the experimental data.

Several factors may contribute to these discrepancies: (i) misalignment in the OCV, (ii) regain of capacity at high SoCs due to electrolyte oxidation \cite{Hartmann2024}, (iii) error in the estimated surface area contributing to continuous SEI growth, (iv) significant electrolyte dry-out caused by extensive SEI formation at 70\% and 100\% SoC, and (v) additional $\textrm{LAM}_\textrm{Si}$ mechanisms at high SoCs, e.g., generation of crystalline silicon \cite{Li2007}. Since all cells are charged to 100\% SoC three times during the CU cycles, the observed increase in $\textrm{LAM}_\textrm{Si}$ at 100\% SoC must occur during the storage period, likely due to slow structural reorganization in partially crystalline silicon.

\begin{figure}[]
	\centering
	\makebox[0pt]{%
    \includegraphics[width=0.41\paperwidth]{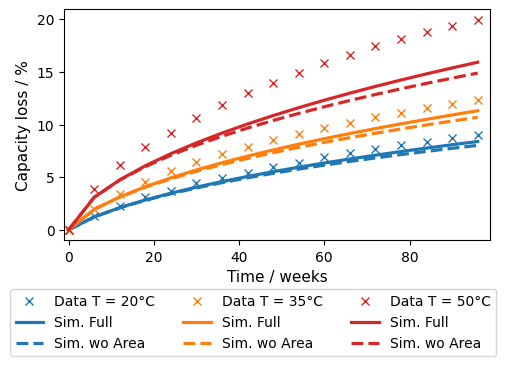}}
	\caption{CL of cells stored at 50\% SoC and varying temperatures ($T = 20$°C, $T = 35$°C, $T = 50$°C) for 96 weeks with CUs every 6 weeks. The dashed lines refer to simulations where the SEI growth on the additional cracked area is neglected. In the main text (and SI Sec I and III) we discuss the used Arrhenius activation energy and higher values.}
	\label{fig.Storage-SoC50-T20-35-50}
\end{figure}

In Fig. \ref{fig.Storage-SoC50-T20-35-50}, we present results for cells stored at 50\% SoC with six-weekly CUs at different temperatures. As expected, CL increases with rising storage temperature. Notably, the model significantly underestimates CL at $T = 50$°C. The simulations that exclude SEI growth on cracked surfaces (dashed lines) yield even lower CL values.

To examine the cause for the deviation, we explored a stronger temperature dependence for the SEI growth parameter $D_\mathrm{e^-}$. Even when assuming a highly exaggerated Arrhenius dependency (see SI Sec I and III), the model still failed to reproduce the experimental CL at $T = 50$°C. We also evaluated the impact of the conservative assumption that SEI growth on additional active areas proceeds at the same rate as on the basic area with an already thicker SEI (see Sec. \ref{subsec:Part.Cracking}). While this assumption slightly underestimates SEI growth, its effect on total CL is minor and not sufficient to explain the observed deviation (see SI Sec. II). Furthermore, an underestimation of crack area cannot account for the discrepancy, as this would contradict other results (see Fig. \ref{fig.Storage-SoC-F6w} and SI Sec. VI). Remaining possible explanations include changes in the SEI composition or passivation behavior at elevated temperatures, or the influence of storage-induced degradation on degradation during subsequent CU cycling.

\subsection{Model Discussion}\label{subsec:limits}

The degradation mechanisms considered in our model are continuous SEI growth and cracking of silicon particles, which leads to $\textrm{LAM}_\textrm{Si}$, particle surface modifications, and instantaneous SEI growth. The model demonstrates strong predictive capability across a wide range of cycling and storage protocols, achieving good agreement with experimental data using a single, consistent parameterization. Nevertheless, during model development, parameterization, and validation, several key aspects and limitations were identified. These are discussed in detail below and further explored in the SI.

In the publication of the experimental data, \citet{Wildfeuer2023} attribute the enhanced degradation observed in storage experiments with frequent CUs to particle cracking - specifically, the generation of additional surface area that accelerates aging during subsequent storage. In contrast, our model explains the high CL at $T = 20$°C under frequent CUs primarily as a result of degradation occurring during the CUs themselves. While the model does account for the increased surface area from cracking and its contribution to accelerated aging in storage, this effect alone is insufficient to fully capture the extent of degradation observed at elevated storage temperatures. The mismatch would indicate an underestimation of the area. However, the area rather seems to be overestimated (see SI Sec. VI). It stands to reason that the high temperature itself triggers additional effects.

The SPMe model generally fails for large current densities ($\geq 4$C), leading to incorrect results and wrong estimation of degradation effects. As a result, the crack growth is overestimated at high discharge currents. We have made efforts to find an all-encompassing parameterization, but the problem remained at very high C-rates ($= 10$C). We expect this to be due to a general limitation of the SPMe model. In the SI, we provide a more detailed analysis of the high C-rate cases and show that, once the stresses and resulting cracks are estimated appropriately, the model can also capture data with high currents. However, this is only possible with a separate parameterization.

Beyond inherent model limitations, several discrepancies between simulation and experiment arise from incomplete cell characterization. The investigated cell was not fully characterized for key parameters such as OCV alignment, electrode-specific overpotentials during cycling, thermal dependencies of battery parameters, and accurate active electrode surface areas. These uncertainties propagate through the model, affecting the prediction of both electrochemical behavior and degradation mechanisms — potentially explaining many of the observed deviations.

In real-world batteries, numerous additional physical phenomena occur that are not accounted for in the current model. While we aim to avoid overfitting by keeping the model simple, some effects are excluded either because they are estimated to have a minor impact (e.g., lithium plating under the conditions studied) or because they cannot be distinguished from existing mechanisms (e.g., whether instantaneous SEI growth occurs within SEI cracks or within cracks in the silicon particle itself). Other neglected effects include the recovery of crystalline silicon phases or phase transitions in silicon that influence volume expansion. These processes may significantly influence degradation but remain outside the model’s scope. Despite these omissions, our model successfully captures the experimental trends across a wide range of cycling and storage scenarios, underscoring its robustness and practical utility for predicting degradation under diverse operating conditions.

\section{Conclusion}\label{sec:Conclusion}

In this work, we developed a physics-based degradation model for batteries with silicon-graphite anodes that aims to describe CL and $\mathrm{SoH}_\mathrm{Si}$ during both calendar and cyclic aging under various usage protocols. A significant challenge in enabling physical conclusions is avoiding overfitting by considering too many complex processes in the battery. To achieve this, the developed degradation model considers only continuous SEI growth and silicon particle cracking due to occurring stresses. These cracks eventually lead to rapid SEI growth on the freshly exposed area and loss of active silicon.

We showed that silicon cracking can play a dominant role in degradation during battery operation, depending on the specific cycling protocol. By investigating different usage profiles, we show that silicon cracking occurs primarily in the low SoC regime, where silicon is preferably used. 
Further, we analyzed cycling cells in the intermediate SoC regime, where we identified SEI growth as the dominant degradation mechanism and validated electron diffusion as a valid SEI growth model for cycling. Compared to cells where the DoD reaches into the low SoC region, overall degradation is significantly lower in the intermediate SoC regime, even at more challenging C-rates.

By applying the model to the storage data, we were able to explain the observed CL for different storage SoCs. Additionally, we were able to explain the observed CL in relation to the number of performed CUs. However, our model does not capture the experimentally observed increase of $\textrm{LAM}_\textrm{Si}$ for storage at high SoC. This deviation suggests another effect for silicon loss, e.g., the formation of crystalline Si phases \cite{Li2007}.

Further, we analyze the model with data at elevated temperatures. For slightly elevated temperatures ($T = 35$°C), we find very good transferability of the model across storage and cycling protocols. However, at very high temperatures ($T = 50$°C), some deviations occur, suggesting that other thermal effects are at play or that the thermal dependencies in the various battery processes are not accurate. These dependencies have a crucial impact on model performance but were outside the scope of this work. For a physical understanding of thermal influence, a separate study is needed to assess the scaling of several effects and the occurrence of additional effects that significantly influence degradation.

To conclude this work, a physics-based degradation model that accounts for SEI growth and silicon cracking describes the observed degradation during storage and cycling at moderate conditions very well. For accurate results under challenging conditions (such as high C-rates and combinations of high temperatures and elevated SoC regions), an underlying, very precisely characterized cell description with correct thermal dependencies is needed. An in-depth cell analysis would be very beneficial for validating the model and mitigating its limitations.

\vspace*{0.5cm}

\section{Author contributions}

\textbf{Micha Philipp}: Conceptualization, Data curation, Formal analysis, Investigation, Methodology, Project administration, Resources, Software, Validation, Visualization, Writing -- original draft, Writing -- review \& editing.
\textbf{Lukas Köbbing}: Methodology, Supervision, Writing -- review \& editing.
\textbf{Alexander Karger}: Investigation, Data curation, Writing -- review \& editing.
\textbf{Andreas Jossen}: Investigation, Data curation, Writing -- review \& editing.
\textbf{Arnulf Latz}: Supervision, Writing -- review \& editing.
\textbf{Birger Horstmann}: Conceptualization, Funding acquisition, Methodology, Project administration, Supervision, Writing -- review \& editing.

\section{Acknowledgements}

This work was supported by the German Aerospace Center (DLR) and the European Union's Horizon Europe research and innovation programme under grant agreement No. 101103997 (DigiBatt) and No. 101104032 (OPINCHARGE). The authors acknowledge support by the state of Baden-Württemberg through bwHPC and the German Research Foundation (DFG) through grant No. INST 40/575-1 FUGG (JUSTUS 2 cluster).

\section{Conflicts of interest}

The authors declare no conflict of interest.


\bibliography{refs}

\end{document}